\documentclass[aps,groupedaddress,showpacs,floatfix]{revtex4}
\usepackage{amssymb}
\usepackage{amsmath}
\usepackage{graphicx}
\DeclareMathOperator{\sgn}{sgn}
\DeclareMathOperator{\Ai}{Ai}
\begin{document}

\title{Two-time free energy distribution function in the KPZ problem}

\author{Victor Dotsenko}

\affiliation{LPTMC, Universit\'e Paris VI, Paris, France}

\affiliation{L.D.\ Landau Institute for Theoretical Physics, Moscow, Russia}

\date{\today}

\begin{abstract}
Following the earlier approach \cite{2time} the explicit expression for the two time free energy
distribution function in one-dimensional random directed polymers is derived  in terms of the Bethe ansatz
replica technique. It is show that such type of the  distribution
function can be represented in the form of a generalized "two-dimensional" Fredholm determinant.

\end{abstract}


\maketitle

\medskip

\section{Introduction}

The problem of directed polymers in a quenched random potential or
equivalent problem of the KPZ-equation
\cite{KPZ} describing the growth in time of an interface
in the presence of noise
have been the subject of intense investigations during past three
decades
\cite{hh_zhang_95,burgers_74,kardar_book,hhf_85,numer1,numer2,kardar_87,bouchaud-orland,Brunet-Derrida,
Johansson,Prahofer-Spohn,Ferrari-Spohn1}.
A few years ago the exact solution for the free energy
probability distribution function (PDF)  has been found
\cite{KPZ-TW1a,KPZ-TW1b,KPZ-TW1c,KPZ-TW2,BA-TW1,BA-TW2,BA-TW3,LeDoussal1,LeDoussal2,goe,LeDoussal3,Corwin,Borodin}.
It was show that this PDF is given by the Tracy-Widom (TW) distribution \cite{TW-GUE}.
The two-point free energy distribution function
which describes joint statistics of the free energies of the directed polymers
coming to two different endpoints has been derived in  \cite{Prolhac-Spohn,2pointPDF,Imamura-Sasamoto-Spohn}.

In all these studies, however, the problems were considered in the so called
"one-time" situation. For the first time the joint probability distribution function
for the free energies of two directed polymers with fixed boundary conditions
at two different times has been studied in the paper \cite{2time} (in terms of non-rigorous replica Bethe
ansatz approach) and in the paper \cite{2time-J} (mathematically rigorous derivation). Unfortunately,
the results obtained in these papers are somewhat inconclusive: the final formulas are expressed in terms rather
complicated mathematical object whose analytic properties are not known. Moreover, for the moment
it is even not clear whether these two results obtained it terms of two different approaches do coincide.

In this paper I am going to derive the joint probability distribution function for somewhat different
two time free energy object. It turns out that compared to the previous calculations \cite{2time,2time-J}
"more symmetric" structure of this object (see below) makes the corresponding calculations much more
simple and the final result can be expressed in the form of a compact formula which could be conditionally
called a "two-dimensional" Fredholm determinant with explicit expression for its kernel,
eqs.(\ref{27})-(\ref{29}).

\vspace{5mm}

The model under consideration is defined in terms of an elastic string $\phi(\tau)$
directed along the $\tau$-axes within an interval $[0,t]$ which passes through a random medium
described by a random potential $V(\phi,\tau)$. The energy of a given polymer's trajectory
$\phi(\tau)$ is
\begin{equation}
   \label{1}
   H[\phi, V] = \int_{0}^{t} d\tau
   \Bigl\{\frac{1}{2} \bigl[\partial_\tau \phi(\tau)\bigr]^2
   + V[\phi(\tau),\tau]\Bigr\};
\end{equation}
where the disorder potential $V[\phi,\tau]$ is Gaussian distributed with a zero mean $\overline{V(\phi,\tau)}=0$
and the $\delta$-correlations ${\overline{V(\phi,\tau)V(\phi',\tau')}} = u \delta(\tau-\tau') \delta(\phi-\phi')$.
The parameter $u$ describes the strength of the disorder.
For the fixed boundary conditions, $\phi(0) = 0, \; \phi(t) = x$, the partition function
of this model is
\begin{equation}
\label{2}
   Z_{t}(x) = \int_{\phi(0)=0}^{\phi(t)=x}
              {\cal D} \phi(\tau)  \;  \mbox{\Large e}^{-\beta H[\phi]}
\; = \; \exp\bigl(-\beta F_{t}(x)\bigr)
\end{equation}
where $\beta$ is the inverse temperature and $F_{t}(x)$ is the free energy.
In the limit $t\to\infty$ the free energy scales as
$\beta F_{t}(x) = \beta f_{0} t + \beta x^{2}/2t + \lambda_{t} f(x)$,
where $f_{0}$ is the selfaveraging free energy density,
$\lambda_{t} = \frac{1}{2}(\beta^{5} u^{2} t)^{1/3}  \propto t^{1/3}$ and $f(x)$
is a random quantity described by the
Tracy-Widom distribution \cite{BA-TW2,BA-TW3,LeDoussal1,LeDoussal2}.
As the first two trivial terms of this free energy can be easily
eliminated by simple redefinition of the partition function, they  will be
omitted in the further calculations, in other words, $ Z_{t}(x) = \exp\bigl(-\lambda_{t} f(x)\bigr)$.

Let us consider two partition functions: the first one, $Z_{t}(0)$, is defined in eq.(\ref{2}) with the zero
boundary conditions,  $\phi(0) = \phi(t) = 0$, while the second one, $\tilde{Z}_{t,\Delta t}(0)$,
in addition to the above zero boundary conditions contains an additional constrain $\phi(t-\Delta t) = 0$.
In other words, in the second case the polymer trajectory before coming to zero at time $t$ is forced
cross the zero at some intermediate time $(t - \Delta t)$. 
\begin{figure}[h]
\begin{center}
   \includegraphics[width=9.0cm]{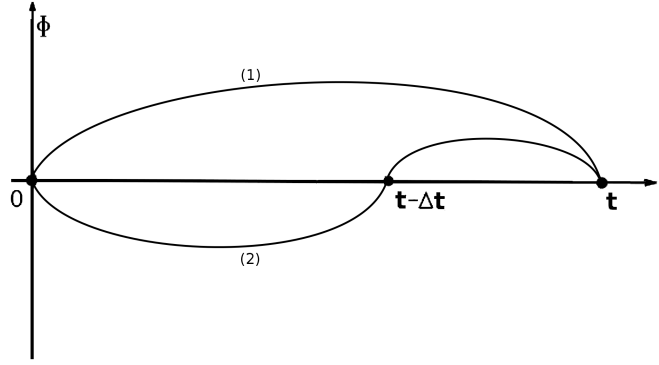}
\caption[]{Schematic representation of two directed polymer paths: (1) with $\phi(0) = \phi(t) = 0$,
and (2) with $\phi(0) = \phi(t-\Delta t) = \phi(t) = 0$}
\end{center}
\label{figure}
\end{figure}

By definition,
\begin{equation}
\label{3}
   Z_{t}(0) = \int_{-\infty}^{+\infty} dx \;
Z_{t-\Delta t}(x)\, Z^{*}_{\Delta t}(x) \; = \; \exp\bigl(-\lambda_{t} f\bigr)
\end{equation}
and
\begin{equation}
\label{4}
   \tilde{Z}_{t,\Delta t}(0) =
Z_{t-\Delta t}(0)\, Z^{*}_{\Delta t}(0) \; = \; \exp\bigl(-\lambda_{t} \tilde{f}\bigr)
\end{equation}
where $Z^{*}_{\Delta t}(x)$ is the partition function of the directed polymer system in which time goes backwards,
from $t$ to $(t - \Delta t)$. For technical reasons (for proper regularization of the integration over $x$ at
$\pm$ infinities) it is convenient to split the partition function
$Z_{t}(0) $ into two parts, the "left" and the "right" ones:
\begin{equation}
 \label{5}
Z_{t}(0) \; = \;
\int_{-\infty}^{0} dx \;
Z_{t-\Delta t}(x)\, Z^{*}_{\Delta t}(x)
\; + \;
\int_{0}^{+\infty} dy \;
Z_{t -\Delta t}(y)\, Z^{*}_{\Delta t}(y)
\end{equation}
In terms of the free energy definitions (\ref{3})-(\ref{4}) one would like to compute the joint probability distribution function
\begin{equation}
\label{6}
W(f_{1}, f_{2}, \Delta) =
\lim_{t\to\infty} \;
\mbox{Prob}\bigl[f > f_{1}; \; \tilde{f} > f_{2} \bigr]
\end{equation}
where it is assumed that in the limit $t\to\infty$ the parameter $\Delta \equiv \Delta t/t$ remains finite.
In terms of the above partition functions (\ref{3})-(\ref{4}) this quantity can be defined as follows:
\begin{equation}
\label{7}
W(f_{1}, f_{2}, \Delta) = \lim_{t\to\infty}
\sum_{N=0}^{\infty} \sum_{K=0}^{\infty}
\frac{(-1)^{N+K}}{N! \; K!}
\exp\bigl(\lambda_{t} N f_{1} + \lambda_{t} K f_{2}\bigr) \;
\overline{Z_{t}^{N}(0) \, \bigl[Z_{t-\Delta t}(0)\, Z^{*}_{\Delta t}(0)\bigr]^{K}}
\end{equation}
where  $\overline{(...)}$ denotes the averaging over the random potential.
Taking into account eq.(\ref{5}) one gets
\begin{eqnarray}
\nonumber
W(f_{1}, f_{2}, \Delta) &=& \lim_{t\to\infty}
\sum_{L,R,K=0}^{\infty}
\frac{(-1)^{L+R+K}}{L!\, R!\, K!}
\exp\Bigl[\lambda_{t} (L+R) f_{1} + \lambda_{t} K f_{2}\Bigr] \;
\int_{-\infty}^{0} dx_{1}...dx_{L} \int_{0}^{+\infty} dy_{R} ... dy_{1} \times
\\
\nonumber
\\
&\times&
\overline{\Biggl(\Bigl[\prod_{a=1}^{L} Z_{t-\Delta t}(x_{a})\Bigr]
                                Z_{t-\Delta t}^{K}(0)
          \Bigl[\prod_{b=1}^{R} Z_{t-\Delta t}(y_{b})\Bigr] \Biggr)\;
          \Biggl(\Bigl[\prod_{a=1}^{L} Z^{*}_{\Delta t}(x_{a})\Bigr]
                                {Z^{*}_{\Delta t}}^{K}(0)
          \Bigl[\prod_{b=1}^{R} Z^{*}_{\Delta t}(y_{b})\Bigr]\Biggr)}
\label{8}
\end{eqnarray}
Introducing:
\begin{equation}
\label{9}
\Psi(x_{1}, ..., x_{N} ; t) \; \equiv \;
\overline{Z_{t}(x_{1}) \, Z_{t}(x_{2}) \, ... \, Z_{t}(x_{N})}
\end{equation}
one can easily show that $\Psi({\bf x}; t)$ is the wave function of $N$-particle
boson system with attractive $\delta$-interaction:
\begin{equation}
   \label{10}
\beta \, \partial_t \Psi({\bf x}; t) =
\frac{1}{2}\sum_{a=1}^{N}\partial_{x_a}^2 \Psi({\bf x}; t)
+\frac{1}{2}\, \kappa \sum_{a\not=b}^{N} \delta(x_a-x_b) \; \Psi({\bf x}; t)
\end{equation}
(where $\kappa = \beta^{3} u$) with the initial condition $\Psi({\bf x}; 0) = \Pi_{a=1}^{N} \delta(x_a)$.
The wave function  $\Psi({\bf x}; t)$ of this quantum problem can be represented in terms of the linear combination
of the corresponding eigenfunctions  of eq.(\ref{10}).
A generic  eigenstate of such system is characterized by $N$ momenta
$\{ Q_{a} \} \; (a=1,...,N)$ which split into
$M$  ($1 \leq M \leq N$) "clusters" each described by
continuous real momenta $q_{\alpha}$ $(\alpha = 1,...,M)$
and characterized by $n_{\alpha}$ discrete imaginary "components"
(for details see \cite{Lieb-Liniger,McGuire,Yang,Calabrese,rev-TW}):
\begin{equation}
   \label{11}
Q_{a} \; \to \; q^{\alpha}_{r} \; = \;
q_{\alpha} - \frac{i\kappa}{2}  (n_{\alpha} + 1 - 2r)
\;\; ; \; \;\;\; \;\;\; \;\;\;
(r = 1, ..., n_{\alpha}\,; \; \; \alpha = 1, ..., M)
\end{equation}
with the global constraint $\sum_{\alpha=1}^{M} n_{\alpha} = N$.
Explicitly,
\begin{equation}
\label{12}
\Psi_{{\bf Q}}({\bf x}) =
\sum_{{\cal P}}  \;
\prod_{1\leq a<b}^{N}
\Biggl[
1 +i \kappa \frac{\sgn(x_{a}-x_{b})}{Q_{{\cal P}_a} - Q_{{\cal P}_b}}
\Biggr] \;
\exp\Bigl[i \sum_{a=1}^{N} Q_{{\cal P}_{a}} x_{a} \Bigr]
\end{equation}
where the vector ${\bf Q}$ denotes the set of all $N$ momenta eq.(\ref{11}) and
the summation goes over $N!$ permutations ${\cal P}$ of $N$ momenta $Q_{a}$,
 over $N$ particles $x_{a}$.
In terms of the above eigenfunctions the solution of eq.(\ref{10}) can be expressed as follows:
\begin{equation}
\label{13}
\Psi({\bf x}; t) = \frac{1}{N!} \int {\cal D}_Q \; |C({\bf Q})|^{2} \; \Psi_{\bf Q}({\bf x}) \Psi^{*}_{\bf Q}(0) \;
\exp\bigl(-t E({\bf Q}) \bigr)
\end{equation}
where
the symbol $\int {\cal D}_Q$ denotes the integration over $M$ continuous parameters $\{ q_{1}, ..., q_{M}\}$,
the summations over $M$ integer parameters $\{ n_{1}, ..., n_{M}\}$ as well as summation over $M=1,..,N$.
$|C({\bf Q})|^{2}$ is the normalization factor,
\begin{eqnarray}
   \nonumber
|C({\bf Q})|^{2} &=& \frac{\kappa^{N}}{\prod_{\alpha=1}^{M}\bigl(\kappa n_{\alpha}\bigr)}
\prod_{\alpha<\beta}^{M}
\frac{\big|q_{\alpha}-q_{\beta} -\frac{i\kappa}{2}(n_{\alpha}-n_{\beta})\big|^{2}}{
      \big|q_{\alpha}-q_{\beta} -\frac{i\kappa}{2}(n_{\alpha}+n_{\beta})\big|^{2}}
\\
\nonumber
\\
&=&
\kappa^{N} \det\Biggl[
   \frac{1}{\frac{1}{2}\kappa n_{\alpha} - i q_{\alpha}
          + \frac{1}{2}\kappa n_{\beta} + iq_{\beta}}\Biggr]_{\alpha,\beta=1,...M}
 \label{14}
\end{eqnarray}
and $E({\bf Q})$ is the eigenvalue (energy) of the eigenstate $\Psi_{\bf Q}({\bf x})$,
\begin{equation}
\label{15}
E({\bf Q}) \; = \;
\frac{1}{2\beta} \sum_{\alpha=1}^{N} Q_{a}^{2} =
 \frac{1}{2\beta} \sum_{\alpha=1}^{M} \; n_{\alpha} q_{\alpha}^{2}
- \frac{\kappa^{2}}{24\beta}\sum_{\alpha=1}^{M} n_{\alpha}^{3}
\end{equation}
In terms of the wave functions (\ref{9})-(\ref{13}) the probability distribution function (\ref{8})
can be expressed as follows:
\begin{eqnarray}
\label{16}
W(f_{1}, f_{2}, \Delta) &=& \lim_{t\to\infty}
\sum_{L,K,R=0}^{\infty}
\frac{(-1)^{L+K+R}}{L!\, K!\, R!} \,
\exp\Bigl[ \lambda_{t} (L+R) f_{1} + \lambda_{t} K f_{2}\Bigr]
\times
\\
\nonumber
\\
\nonumber
&\times&
\int_{-\infty}^{0} dx_{1}...dx_{L} \int_{0}^{+\infty} dy_{R} ... dy_{1}
\Psi\bigl(x_{1},...,x_{L}, \underbrace{0, ..., 0}_{K}, y_{R}, ..., y_{1} ; \, (t - \Delta t) \bigr) \;
\Psi^{*}\bigl(x_{1},...,x_{L}, \underbrace{0, ..., 0}_{K}, y_{R}, ..., y_{1} ; \, \Delta t\bigr)
\end{eqnarray}
where the second (conjugate) wave function represent the "backward" propagation
from the time moment $t$ to the previous time moment $t-\Delta t$.
Schematically the above expression is represented in Figure 1.
\begin{figure}[h]
\begin{center}
   \includegraphics[width=10.0cm]{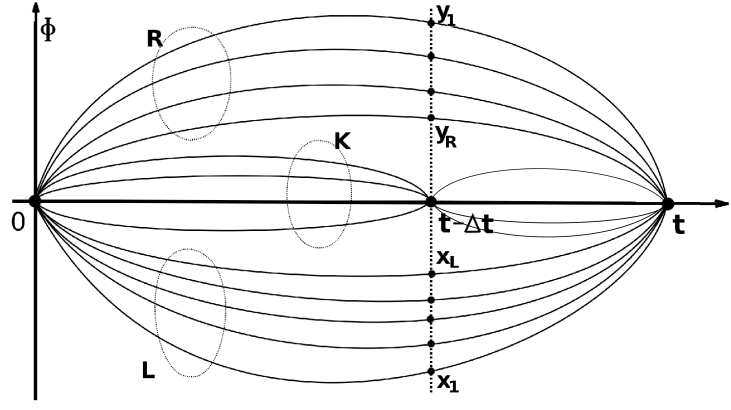}
\caption[]{Schematic representation of the directed polymer paths
corresponding to eq.(\ref{16})}
\end{center}
\label{figure1}
\end{figure}
The above expression is quite similar to eq.(23) in \cite{2time} although here its structure is "more symmetric"
which essentially simplify further calculations. Repeating the same steps as in the paper \cite{2time},
in the limit $t\to\infty$ one eventually gets (sf eq.(65) in  \cite{2time}):
\begin{eqnarray}
 \nonumber
W(f_{1},f_{2},\Delta) &=&
1 + \sum_{M=1}^{\infty}
\frac{(-1)^{M}}{(M!)^{2}}
\\
\nonumber
\\
\nonumber
&\times&
\prod_{\alpha=1}^{M}
\Biggl[
\int\int_{-\infty}^{+\infty} \frac{dq_{\alpha}dp_{\alpha}}{(2\pi)^{2}}
\int_{-\infty}^{+\infty} dy \;
\Ai\bigl(y + (1-\Delta) q_{\alpha}^{2} + \Delta p_{\alpha}^{2}- f_{1}\bigr)
\times
\\
\nonumber
\\
\nonumber
&\times&
\int_{{\cal C}} d{z_{1}}_{\alpha} d{z_{2}}_{\alpha} d{z_{3}}_{\alpha}
\Bigl(\frac{1}{{z_{1}}_{\alpha} {z_{2}}_{\alpha} {z_{3}}_{\alpha}}
- \delta({z_{1}}_{\alpha}) \delta({z_{2}}_{\alpha})\delta({z_{3}}_{\alpha}) \Bigr)
J\Bigl(q_{\alpha}- p_{\alpha}, {z_{1}}_{\alpha}, {z_{2}}_{\alpha}, {z_{3}}_{\alpha}\Bigr)
\times
\\
\nonumber
\\
\nonumber
&\times&
\exp\bigl\{ ({z_{1}}_{\alpha} + {z_{2}}_{\alpha} + {z_{3}}_{\alpha}) y + {z_{3}}_{\alpha} (f_{2} - f_{1})\bigr\}
\Biggr]
\times
\\
\nonumber
\\
\nonumber
&\times&
\det\Biggl[
\frac{1}{
{z_{1}}_{\alpha} + {z_{2}}_{\alpha} + {z_{3}}_{\alpha}-iq_{\alpha} +
{z_{1}}_{\alpha'} + {z_{2}}_{\alpha'} + {z_{3}}_{\alpha'}+iq_{\alpha'}}
\Biggr]_{\alpha,\alpha'=1,...,M}
\times
\\
\nonumber
\\
&\times&
\det\Biggl[
\frac{1}{
{z_{1}}_{\alpha} + {z_{2}}_{\alpha} + {z_{3}}_{\alpha}+ip_{\alpha} +
{z_{1}}_{\alpha'} + {z_{2}}_{\alpha'} + {z_{3}}_{\alpha'}-ip_{\alpha'}}
\Biggr]_{\alpha,\alpha'=1,...,M}
\label{17}
\end{eqnarray}
where the integration contour ${\cal C} $ is shown in figure 2, and
\begin{equation}
\label{18}
J\Bigl(q- p, {z_{1}}, {z_{2}}, {z_{3}}\Bigr) \; = \;
\Biggl(1 + \frac{z_{1}}{z_{2} +  z_{3} + \frac{1}{2} i (q-p)^{(-)}}\Biggr)
\Biggl(1 + \frac{z_{2}}{z_{1} +  z_{3} - \frac{1}{2} i (q-p)^{(+)}}\Biggr)
\end{equation}
Here $(q-p)^{(\pm)} \equiv (p-q) \pm i \epsilon$ where the parameter $\epsilon$ has to be set to zero
at the end.
\begin{figure}[h]
\begin{center}
   \includegraphics[width=6.0cm]{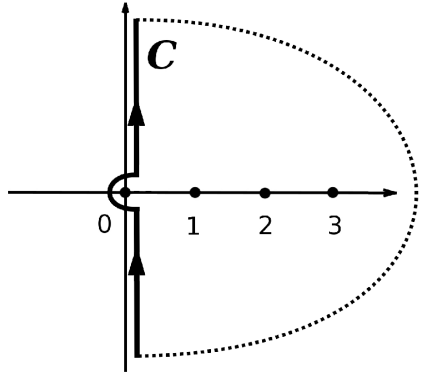}
\caption[]{ Contour of integration ${\cal C}$ in eq.(\ref{17}).}
\end{center}
\label{figure2}
\end{figure}
Determinants in eq.(\ref{17}) can be represented in the following integral form:
\begin{eqnarray}
 \nonumber
\det\Biggl[
\frac{1}{
z_{\alpha} - iq_{\alpha} + z_{\beta} + iq_{\beta}}
\Biggr]_{\alpha,\beta=1,...,M}
&=&
\sum_{{\cal P}\in S_{M}} (-1)^{\bigl[{\cal P}\bigr]}
\prod_{\alpha=1}^{M}
\bigl( z_{\alpha} - iq_{\alpha} + z_{{\cal P}_\alpha} + iq_{{\cal P}_\alpha} \bigr)^{-1}
\\
\nonumber
\\
\nonumber
&=&
\prod_{\alpha=1}^{M}
\Biggl[\int_{0}^{\infty} du_{\alpha}\Biggr]
\sum_{{\cal P}\in S_{M}} (-1)^{\bigl[{\cal P}\bigr]}
\prod_{\alpha=1}^{M}
\exp
\Bigl\{
- \bigl(z_{\alpha} + z_{{\cal P}_\alpha}\bigr) u_{\alpha}
+i \bigl(q_{\alpha} - q_{{\cal P}_\alpha}\bigr) u_{\alpha}
\Bigr\}
\\
\nonumber
\\
\nonumber
&=&
\prod_{\alpha=1}^{M}
\Biggl[
\int_{0}^{\infty} du_{\alpha}
\Biggr]
\sum_{{\cal P}\in S_{M}} (-1)^{\bigl[{\cal P}\bigr]}
\prod_{\alpha=1}^{M}
\exp
\Bigl\{
- z_{\alpha} \bigl(u_{\alpha} + u_{{\cal P}_\alpha}\bigr)
+ i q_{\alpha} \bigl(u_{\alpha} - u_{{\cal P}_\alpha}\bigr)
\Bigr\}
\\
\label{19}
\end{eqnarray}
Substituting this representation into eq.(\ref{17}) and performing integrations over
$z_{1}, z_{2}$ and $z_{3}$ one obtains the following result
\begin{eqnarray}
 \nonumber
W(f_{1},f_{2},\Delta) &=&
1 + \sum_{M=1}^{\infty}
\frac{(-1)^{M}}{(M!)^{2}}
\sum_{{\cal P},{\cal P}'\in S_{M}} (-1)^{\bigl[{\cal P}\bigr]+\bigl[{\cal P}'\bigr]}
\prod_{\alpha=1}^{M}
\Biggl[
\int\int_{0}^{\infty} du_{\alpha}dv_{\alpha}
\Biggr]
\\
\nonumber
\\
\nonumber
&\times&
\prod_{\alpha=1}^{M}
\Biggl[
\int\int_{-\infty}^{+\infty} \frac{dq_{\alpha}dp_{\alpha}}{(2\pi)^{2}}
\int_{-\infty}^{+\infty} dy \;
\Ai\Bigl(y + (1-\Delta) q_{\alpha}^{2} + \Delta p_{\alpha}^{2}- f_{1}
+ u_{\alpha} + v_{\alpha} + u_{{\cal P}_\alpha} + v_{{\cal P'}_\alpha} \Bigr)
\\
\nonumber
\\
&\times&
\exp
\Bigl\{
+ i q_{\alpha} \bigl(u_{\alpha} - u_{{\cal P}_\alpha}\bigr)
- i p_{\alpha} \bigl(v_{\alpha} - v_{{\cal P'}_\alpha}\bigr)
\Bigr\} \;
S\bigl[ (q_{\alpha} - p_{\alpha}); \; y; \; (f_{2}-f_{1})\bigr]
\Biggr]
\label{20}
\end{eqnarray}
where
\begin{eqnarray}
\nonumber
S\bigl[ (q - p); \; y; \; (f_{2}-f_{1})\bigr]  &=&
4\pi \delta(y) \delta(q-p) \; \theta(f_{1} - f_{2}) \; +
\\
\nonumber
\\
&+&
\Biggl[
\theta(f_{2} - f_{1} + y) \theta(f_{2} - f_{1} - y)  \; - \;
4 \delta(y) \frac{\sin\bigl[\frac{1}{2}(q-p)(f_{2} - f_{1})\bigr]}{(q-p)}
\Biggr]
 \, \theta(f_{2} - f_{1})
 \label{21}
\end{eqnarray}
According to the above equation in the cases: (a) $f_{2} < f_{1}$ and (b) $f_{2} > f_{1}$,
one finds two essentially different results for the probability distribution function $W(f_{1},f_{2},\Delta)$.

\vspace{3mm}

{\bf (a)} $ \boldsymbol{f_{2} < f_{1}}$. In this case the distribution function $W(f_{1},f_{2},\Delta)$ turns out to be
independent of $f_{2}$ and $\Delta$:
\begin{eqnarray}
 \nonumber
W\bigl(f_{1},f_{2}, \Delta\bigr)\Big|_{f_{2} < f_{1}} &\equiv& 
W(f_{1}) =
1 + \sum_{M=1}^{\infty}
\frac{(-1)^{M}}{(M!)^{2}}
\sum_{{\cal P},{\cal P}'\in S_{M}} (-1)^{\bigl[{\cal P}\bigr]+\bigl[{\cal P}'\bigr]}
\prod_{\alpha=1}^{M}
\Biggl[
\int\int_{0}^{\infty} du_{\alpha}dv_{\alpha}
\Biggr]
\\
\nonumber
\\
\nonumber
&\times&
\prod_{\alpha=1}^{M}
\Biggl[
\int_{-\infty}^{+\infty} \frac{dq_{\alpha}}{\pi}
\Ai\Bigl(q_{\alpha}^{2} - f_{1}
+ (u_{\alpha}+v_{{\cal P'}_\alpha}) + (v_{\alpha} + u_{{\cal P}_\alpha} ) \Bigr)
\exp
\Bigl\{
i q_{\alpha} \bigl[(u_{\alpha}+v_{{\cal P'}_\alpha}) - (v_{\alpha} + u_{{\cal P}_\alpha} )\bigr]
\Bigr\}
\Biggr]
\\
\nonumber
\\
\nonumber
\\
\nonumber
&=&
1 + \sum_{M=1}^{\infty}
\frac{(-1)^{M}}{(M!)^{2}}
\sum_{{\cal P},{\cal P}'\in S_{M}} (-1)^{\bigl[{\cal P}\bigr]+\bigl[{\cal P}'\bigr]}
\prod_{\alpha=1}^{M}
\Biggl[
\int\int_{0}^{\infty} du_{\alpha}dv_{\alpha}
\Biggr]
\\
\nonumber
\\
&\times&
\prod_{\alpha=1}^{M}
\Biggl[
2^{2/3}
\Ai\bigl[2^{1/3}(u_{\alpha}+v_{{\cal P'}_\alpha} - f_{1}/2))\bigr] \;
\Ai\bigl[2^{1/3}(u_{{\cal P}_\alpha} + v_{\alpha} - f_{1}/2))\bigr]
\Biggr]
\label{22}
\end{eqnarray}
Redefining $u_{\alpha} \to 2^{-1/3} u_{\alpha}$,
$v_{\alpha} \to 2^{-1/3} v_{\alpha}$, and taking into account that
\begin{equation}
 \label{23}
\prod_{\alpha=1}^{M}
\Ai\Bigl[u_{{\cal P}_{\alpha}} + v_{\alpha} - f_{1}/2^{2/3}\Bigr] \; = \;
\prod_{\alpha=1}^{M}
\Ai\Bigl[u_{\alpha} + v_{{\cal P}^{-1}_{\alpha}} - f_{1}/2^{2/3}\Bigr]
\end{equation}
we obtain
\begin{equation}
W(f_{1}) =
1 + \sum_{M=1}^{\infty}
\frac{(-1)^{M}}{(M!)^{2}}
\prod_{\alpha=1}^{M}
\Biggl[
\int_{0}^{\infty} dv_{\alpha}
\Biggr]
\sum_{{\cal P},{\cal P}'\in S_{M}} (-1)^{\bigl[{\cal P}\bigr]+\bigl[{\cal P}'\bigr]}
\prod_{\alpha=1}^{M}
\Biggl[
 K\bigl(v_{{\cal P}^{-1}_{\alpha}} - f_{1}/2^{2/3}; \;
v_{{\cal P'}_{\alpha}} - f_{1}/2^{2/3} \bigr)
\Biggr]
\label{24}
\end{equation}
where
\begin{equation}
\label{25}
K\bigl(v; \; v' \bigr) \; = \; \int_{0}^{\infty} du \;
\Ai(u + v) \Ai(u + v')
\end{equation}
is the Airy kernel. Redefining the permutations, ${\cal P} + {\cal P'} \; \to  \; {\cal P}$,
we eventually get
\begin{equation}
W(f_{1}) =
1 + \sum_{M=1}^{\infty}
\frac{(-1)^{M}}{M!}
\prod_{\alpha=1}^{M}
\Biggl[
\int_{0}^{\infty} dv_{\alpha}
\Biggr]
\sum_{{\cal P}\in S_{M}} (-1)^{\bigl[{\cal P}\bigr]}
\prod_{\alpha=1}^{M}
\Biggl[
 K\bigl(v_{\alpha} - f_{1}/2^{2/3}; \;
v_{{\cal P}_{\alpha}} - f_{1}/2^{2/3} \bigr)
\Biggr] \;  = \; F_{2}\bigl(- f_{1}/2^{2/3} \bigr)
\label{26}
\end{equation}
Thus we have got the usual Tracy-Widom distribution for the free energy
$f_{1}$ of the directed polymer with the zero boundary conditions, as
it should be (it is evident that in the limit $t\to\infty$ with the probability one
the free energy ($f_{2}$) of the polymer which is forced to pass par the zero at some intermediate time
$(t - \Delta t)$ is larger than the free energy ($f_{1}$) of the polymer which is free of this
condition).

\vspace{3mm}

{\bf (b)} $ \boldsymbol{f_{2} > f_{1}}$. In this case the situation becomes more tricky. According to eqs.(\ref{20}) and (\ref{21}) we get:
\begin{equation}
W\bigl(f_{1},f_{2}, \Delta\bigr)\Big|_{f_{2} > f_{1}} =
1 + \sum_{M=1}^{\infty}
\frac{(-1)^{M}}{(M!)^{2}}
\prod_{\alpha=1}^{M}
\Biggl[
\int\int_{0}^{\infty} du_{\alpha} dv_{\alpha}
\Biggr]
\sum_{{\cal P},{\cal P}'\in S_{M}} (-1)^{\bigl[{\cal P}\bigr]+\bigl[{\cal P}'\bigr]}
\prod_{\alpha=1}^{M}
\Biggl[
 K_{(u_{\alpha}, v_{\alpha}); \; (u_{{\cal P}_{\alpha}}, v_{{\cal P'}_{\alpha}})}\bigl(f_{1},f_{2},\Delta\bigr)
\Biggr]
\label{27}
\end{equation}
where
\begin{eqnarray}
\nonumber
 K_{(u, v); \; (u', v')}\bigl(f_{1},f_{2},\Delta\bigr) &=&
\int_{-(f_{2}-f_{1})}^{(f_{2}-f_{1})} dy \;
\int\int_{-\infty}^{+\infty} \frac{dq dp}{(2\pi)^{2}}
\Ai\bigl[y - f_{1} + (1-\Delta)q^{2} + \Delta p^{2} + u + v + u' + v'\bigr] \, \times
\\
\nonumber
\\
\nonumber
&\times&
\exp\bigl\{iq(u-u') - ip(v-v')\bigr\} \; -
\\
\nonumber
\\
\nonumber
&-4&
\int\int_{-\infty}^{+\infty} \frac{dq dp}{(2\pi)^{2}} \;
\frac{\sin\Bigl(\frac{1}{2}(q-p)(f_{2}-f_{1})\Bigr)}{q-p} \,
\Ai\bigl[- f_{1} + (1-\Delta)q^{2} + \Delta p^{2} + u + v + u' + v'\bigr] \, \times
\\
\nonumber
\\
&\times&
\exp\bigl\{iq(u-u') - ip(v-v')\bigr\}
\label{28}
\end{eqnarray}
The structure in eq.(\ref{27}) could be conditionally represented as a
"two-dimensional" Fredholm determinant:
\begin{equation}
\label{29}
W\bigl(f_{1},f_{2}, \Delta\bigr)\Big|_{f_{2} > f_{1}} \; = \;
\boldsymbol{\det}\bigl[\hat{\boldsymbol{1}} \; - \; \hat{\boldsymbol{K}} \bigr]
\end{equation}
where $\hat{\boldsymbol{K}} \; = \; K_{\boldsymbol{\xi}, \boldsymbol{\xi'}}\bigl(f_{1},f_{2},\Delta\bigr)$
is the integral operator function of three parameters
$f_{1},f_{2}$ and $\Delta$ on the two-dimensional space $\boldsymbol{\xi} \equiv (u,v)$
with $(u, v) \in [0, \infty)$ and with the kernel given in eq.(\ref{28}).

Eqs.(\ref{27})-(\ref{29}) constitute the central result of this work for the two-time free energy
probability distribution function of two directed polymers with the two types of the zero boundary conditions:
$\phi(0) = \phi(t) = 0$ and $\phi(0) = \phi(t-\Delta t) = \phi(t) = 0$ in the thermodynamic limit
$t \to \infty$ such the parameter $\Delta \equiv \Delta t/t$ remains finite.
Analytic properties of the mathematical object defined in Eqs.(\ref{27})-(\ref{28}) remains to be
investigated.

\acknowledgments

An essential part of this work was done during the workshop "Stochastic processes in random media", 
held at the Institute for Mathematical Sciences, National University of Singapore. 
The author wishes to thank the IMS for warm hospitality and financial support, 
and also acknowledges a partial financial support from the  ONRG Grant N62909-15-1-C076.
The author also acknowledge hospitality and support from Galileo Galilei Institute, 
and from the scientific program on "Statistical Mechanics, Integrability and Combinatorics" (Florence, 11 May - 3 July, 2015).

\end{document}